\documentclass[twocolumn,floatfix,a4paper,showpacs,showkeys,nofootinbib,reprint,prl]{revtex4-1}
\usepackage{epsfig}
\usepackage{latexsym}
\usepackage{xspace}
\usepackage[colorlinks=true,linktocpage=true,linkcolor=blue,citecolor=blue,allcolors=blue]{hyperref}
\usepackage{url}
\usepackage[utf8]{inputenc}
\usepackage{indentfirst}
\usepackage{enumerate}
\usepackage{color}

\usepackage[caption=false,position=top]{subfig}

\usepackage{amsmath}
\usepackage{amssymb}

\usepackage[english]{babel}
\usepackage{url}

\AtBeginDocument{\renewcommand{\natexlab}[1]{#1}}

\newcommand{\mean}[1]{\langle #1 \rangle}

\newcommand{\sNN}{\sqrt{s_{\rm NN}}}

\newcommand{\eq}[1]{\begin{align} #1 \end{align}}

\begin{document}

\title{
Thermodynamic approach to proton number fluctuations in baryon-rich heavy-ion matter created at moderate collision energies
}

\author{Volodymyr Vovchenko}
\affiliation{Nuclear Science Division, Lawrence Berkeley National Laboratory, 1 Cyclotron Road, Berkeley, CA 94720, USA}
\affiliation{Institute for Nuclear Theory, University of Washington, Box 351550, Seattle, WA 98195, USA}
\affiliation{Frankfurt Institute for Advanced Studies, Giersch Science Center,
D-60438 Frankfurt am Main, Germany}

\author{Volker Koch}
\affiliation{Nuclear Science Division, Lawrence Berkeley National Laboratory, 1 Cyclotron Road, Berkeley, CA 94720, USA}

\begin{abstract}
We develop a framework to relate proton number cumulants measured in heavy-ion collisions within a momentum space acceptance to the susceptibilities of baryon number, assuming that particles are emitted from a fireball with uniform distribution of temperature and baryochemical potential, superimposed on a hydrodynamic flow velocity profile.
The rapidity acceptance dependence of proton cumulants measured by the HADES Collaboration in $\sqrt{s_{\rm NN}} = 2.4$~GeV Au-Au appears to be consistent with thermal emission of nucleons from a grand-canonical heat bath, with the extracted baryon number susceptibilities exhibiting an hierarchy 
$\chi_4^B \gg -\chi_3^B \gg \chi_2^B \gg \chi_1^B$.
Naively, this could indicate large non-Gaussian fluctuations that might point to the presence of the QCD critical point close to the chemical freeze-out at $T \sim 70$~MeV and $\mu_B \sim 850-900$~MeV.
However, the description of the experimental data at large rapidity acceptances becomes challenging once the effect of exact baryon number conservation is incorporated, suggesting that more theoretical and experimental studies are needed to reach a firm conclusion.
\end{abstract}

\maketitle

\paragraph{\bf Introduction.}

The search for the QCD critical point in heavy-ion collisions is centered on the measurements of event-by-event fluctuations of (net) proton number~\cite{Hatta:2003wn,Bzdak:2019pkr}.
The corresponding measurements have been performed by a number of experiments, including the ALICE experiment at the LHC~\cite{ALICE:2019nbs}, the NA61/SHINE experiment at SPS~\cite{Gazdzicki:2015ska}, and the STAR experiment at RHIC~\cite{STAR:2020tga,STAR:2021iop}.
At $\sNN \gtrsim 20$~GeV these measurements are consistent with being driven by noncritical physics such as baryon number conservation and repulsive interactions~\cite{Braun-Munzinger:2020jbk,Vovchenko:2021kxx}. 
At lower energies, the data indicate an excess of two-proton correlations over a non-critical reference, with deviations increasing as the collision energy is decreased~\cite{Vovchenko:2021kxx}. 
Recent data at $\sNN = 3$~GeV from the STAR fixed target program~\cite{STAR:2021fge} are consistent with this trend.
One potential source of these excess correlations among protons could be the presence of the critical point.

Recently, the HADES Collaboration has presented measurements of the four leading proton number cumulants and integrated correlation functions~\cite{HADES:2020wpc} measured in Au-Au collisions at $\sNN = 2.4$~GeV.
The data, which are corrected for volume fluctuations, show strong dependence on the phase-space acceptance, approaching the Poisson limit for narrow rapidity bins and indicating the presence of long-range multi-particle correlations in rapidity as the acceptance is expanded.
In this work we analyze the HADES data in the context of thermal emission of particles from an interacting fireball.
We show that, in this scenario, the measurements in momentum space can be unfolded to extract the grand-canonical baryon number susceptibilities of QCD matter probed by the experiment and provide information on the phase structure of baryon-rich QCD matter.
In particular, we discuss the possible presence of the QCD critical point in the baryon-rich region of QCD matter as a driving mechanism behind the behavior of experimental data as well as effect of baryon number conservation.

\paragraph{\bf Cumulants and correlation functions of protons emitted from an expanding fireball.}

The basis for describing the emission of particles from an expanding fireball is the Cooper-Frye formula~\cite{Cooper:1974mv}, which determines their invariant momentum spectrum
\eq{\label{eq:CF}
\omega_p \frac{d N}{d^3 p} = \int_{\sigma(x)} d \sigma_\mu (x) \, p^\mu \, f[u^\nu(x) p_\nu].
}
Here $f \propto \exp\left[ -u^\nu(x) p_\nu/T(x) \right]$ is the phase-space distribution of particles in thermodynamic equilibrium, $p^\mu$ is their four-momentum\footnote{We neglect the effects of Fermi-Dirac statistics on the momentum distributions in the present work.}, and $d \sigma_\mu (x)$ is the invariant element of the freeze-out hypersurface from which the particles are emitted.
In Ref.~\cite{Vovchenko:2021kxx} the Cooper-Frye procedure was generalized to describe not only the first moments of particle distributions in a given momentum acceptance, but also the cumulants of their multiplicity distribution that are connected to the grand-canonical (baryon number) susceptibilities.
The generalized formalism was used to calculate the proton number cumulants and correlation functions in Au-Au collisions at $\sNN = 7.7-200$~GeV within numerical relativistic hydrodynamics simulations, as appropriate to the beam energy scan program at RHIC.
Here we extend these considerations to lower collision energies, with specific applications to $\sNN = 2.4$~GeV Au-Au collisions studied by the HADES experiment at GSI-SIS.

Compared to higher collision energies, several modifications and simplifications can be made.
First, at $\sNN = 2.4$~GeV one can neglect the production of antiparticles, as well as strange particles.
Second, instead of using the numerical output from hydro simulations, it is possible to employ a single freeze-out scenario parameterized by an appropriate blast-wave-like hypersurface.
Indeed, as recently shown in Ref.~\cite{Harabasz:2020sei}, the $p_T$ spectra of light flavored hadrons in central Au-Au collisions at HADES can be reasonably well described through a spherically-symmetric Siemens-Rasmussen blast-wave model with a Hubble-like radial flow.
In this case the freeze-out takes place suddenly at a constant Cartesian time $t = \rm const$ and within a spherical spatial volume of radius $R$.
The hypersurface element reads $d \sigma_\mu = (1,0,0,0) \, r^2 \sin \theta \, d \theta \, d \phi \, dr$, where $0 < \theta < \pi$, $0 < \phi < 2\pi$, and $0 < r < R$.
The hydrodynamic flow is spherically symmetric and Hubble-like,
$u^{\mu} = \gamma(r) \, (1,v(r) \mathbf{e}_r)$ with $\gamma(r) = (1-v^2(r))^{-1/2}$ and $v(r) = \tanh(Hr)$.

An important simplification of the single freeze-out scenario is that all elements of the freeze-out hypersurface are characterized by constant values of the temperature and baryochemical potential. 
This fact is used here to show how the measured fluctuations of baryon~(or proton) number are related to the values of baryon number susceptibilities $\chi_n^B$ at the freeze-out point in the QCD phase diagram.
Indeed, let us first start with the grand canonical limit, where the emission of baryons from each hypersurface element $d \sigma(x)$ proceeds independently.
The cumulants $d \kappa_n^B (x)$ of baryon distribution emitted from $d \sigma(x)$ is determined by the susceptibilities $\chi_n^B$, namely
\eq{
d \kappa_n^B (x) & = dV^{\rm eff} (x) \, T^3 \, \chi_n^B~.
}
Here $dV^{\rm eff} (x) = d \sigma_\mu(x) u^{\mu}(x)$, thus
\eq{\label{eq:dkappa}
d \kappa_n^B (x) & = T^3 \, \chi_n^B \, \gamma(r) \, r^2 \, \sin \theta \, d \theta \, d \phi \, d r~.
}
The cumulants of the baryon number distribution in the full phase-space in the grand-canonical limit are obtained by integrating Eq.~\eqref{eq:dkappa} over the freeze-out hypersurface, i.e. $\kappa_n^B = \int_{\sigma(x)} d \kappa_n^B (x) = V^{\rm eff} T^3 \chi_n^B$.
The experiment measures particles in a restricted part of the full phase-space, however.
Let us denote the momentum acceptance by $\Delta p_{\rm acc}$.
Every nucleon emitted from $d \sigma(x)$ ends up in $\Delta p_{\rm acc}$ with a certain probability $p_{\rm acc}(x;\Delta p_{\rm acc})$ which can be evaluated through the Cooper-Frye formula:
\eq{\label{eq:pacc}
p_{\rm acc}(x;\Delta p_{\rm acc}) = \frac{\int_{p \in \Delta p_{\rm acc}} \frac{d^3 p}{\omega_p} \, d\sigma_\mu(x) p^\mu f[u^\nu(x)p_\nu]}{\int_{p } \frac{d^3 p}{\omega_p} \, d\sigma_\mu(x) p^\mu f[u^\nu(x)p_\nu]}~.
}
The momentum acceptance can thus be modeled by convoluting $d \kappa_n^B (x)$ with the binomial distribution~\cite{Kitazawa:2012at}, yielding~(see Ref.~\cite{Vovchenko:2021kxx} for details):
\eq{
d \kappa_n^B (x; \Delta p_{\rm acc}) = \sum_{l=1}^n \, d \kappa_l^{B} (x) \,
B_{n,l}\left(\phi'_t, \ldots, \phi^{(n-l+1)}_t \right)~.
}
Here $\phi \equiv \phi[t,p_{\rm acc}] = \ln(1- p_{\rm acc}+e^t p_{\rm acc})$ with $p_{\rm acc} \equiv p_{\rm acc}(x;\Delta p_{\rm acc})$, $\phi^{(n)}_t \equiv \left. \frac{d^n \phi}{d t^n}  \right|_{t=0}$, and $B_{n,l}$ are partial Bell polynomials.
As a result, the total cumulants of accepted baryons $\kappa_n^B (\Delta p_{\rm acc}) = \int_x d \kappa_n^B (x; \Delta p_{\rm acc})$ can be expressed as linear combinations of $\chi_n^B$:
\eq{\label{eq:kappaBacc}
\kappa_n^B (\Delta p_{\rm acc}) =  \sum_{l=1}^n \alpha_{n,l} (\Delta p_{\rm acc}) \, T^3 \, \chi_l^B
}
with
\eq{\label{eq:alpha}
\alpha_{n,l} (\Delta p_{\rm acc}) = \int_{\sigma(x)} dV^{\rm eff}(x) \, B_{n,l}\left(\phi'_t, \ldots, \phi^{(n-l+1)}_t \right)~.
}
Note that these calculations automatically incorporate the effect of thermal broadening~\cite{Ohnishi:2016bdf}, through the integration over the Maxwell-Boltzmann momentum distribution $f$ in Eq.~\eqref{eq:pacc}.

The experimentally measured protons constitute a fraction of all baryons.
Based on the isospin content of the colliding nuclei, one can estimate that roughly $q^{\rm iso} \approx 0.4$ of all baryons are protons.
Furthermore, at low collision energies, a significant fraction $q^{\rm nucl}$ of protons becomes bound into light nuclei, with $q^{\rm nucl} \approx 0.375$ based on HADES measurements of proton and light nuclei production~\cite{Harabasz:2020sei}. 
Assuming that this process happens sometime after the chemical freeze-out, for instance via coalescence, this implies that a fraction $q = q^{\rm iso} (1 - q^{\rm nucl}) \approx 0.25$ of all accepted baryons is ultimately measured in the experiment.
To leading order, this effect can be modeled by convoluting $\kappa_n^B (\Delta p_{\rm acc})$ with a binomial distribution with acceptance $q$.
As a result, the cumulants $\kappa_n^p (\Delta p_{\rm acc})$ of \emph{measured} protons can be expressed as linear combinations of the grand-canonical baryon susceptibilities $\chi_n^B$
\eq{\label{eq:kappanp}
\kappa_n^p (\Delta p_{\rm acc}) =  \sum_{l=1}^n \tilde{\alpha}_{n,l}(\Delta p_{\rm acc}) \, T^3 \, \chi_l^B
}
with
\eq{\label{eq:alphatilde}
\tilde{\alpha}_{n,l} (\Delta p_{\rm acc}) = \sum_{k=1}^n \alpha_{k,l} (\Delta p_{\rm acc}) \, B_{n,k}\left(\gamma'_t, \ldots, \gamma^{(n-k+1)}_t \right)~.
}
Here $\gamma(t,q) = \ln(1- q+e^t q)$ and $\gamma^{(n)}_t \equiv \left. \frac{d^n \gamma}{d t^n}  \right|_{t=0}$.

Equation~\eqref{eq:kappanp} shows that cumulants $\kappa_n^p (\Delta p_{\rm acc})$ measured in a particular acceptance $\Delta p_{\rm acc}$ are expressed as linear combinations of baryon number susceptibilities $\chi_l^B$.
As the effect of exact baryon conservation has not been taken into account, this approach should work only when this effect can be neglected, i.e. for a sufficiently small acceptance $\Delta y_{\rm acc} \ll \Delta y_{\rm beam}$~\cite{Vovchenko:2020tsr}. 
Below, we use this fact to extract $\chi_l^B$ from the experimental data of the HADES Collaboration.

\paragraph{\bf Analysis of the HADES data.}

We analyze the four leading proton cumulants in 0-5\% Au-Au collisions at $\sNN = 2.4$~GeV measured by the HADES Collaboration in Ref.~\cite{HADES:2020wpc}.
To fix the parameters of the freeze-out hypersurface we utilize the results of Refs.~\cite{Harabasz:2020sei,Motornenko:2021nds,HarabaszTalk}, where the chemical and kinetic freeze-out conditions were studied.
As shown in~\cite{Motornenko:2021nds}, a consistent description of hadron yields is obtained for chemical freeze-out temperature of $T \approx 70$~MeV.
In~\cite{Harabasz:2020sei,HarabaszTalk} it was shown that the momentum spectra can be reasonably well described within a single freeze-out scenario. We take the following parameters based on Ref.~\cite{HarabaszTalk}: $T = 70$~MeV, $R = 6.1$~fm, $H = 0.097$~fm$^{-1}$.

\begin{figure}[t]
  \centering
  \includegraphics[width=.41\textwidth]{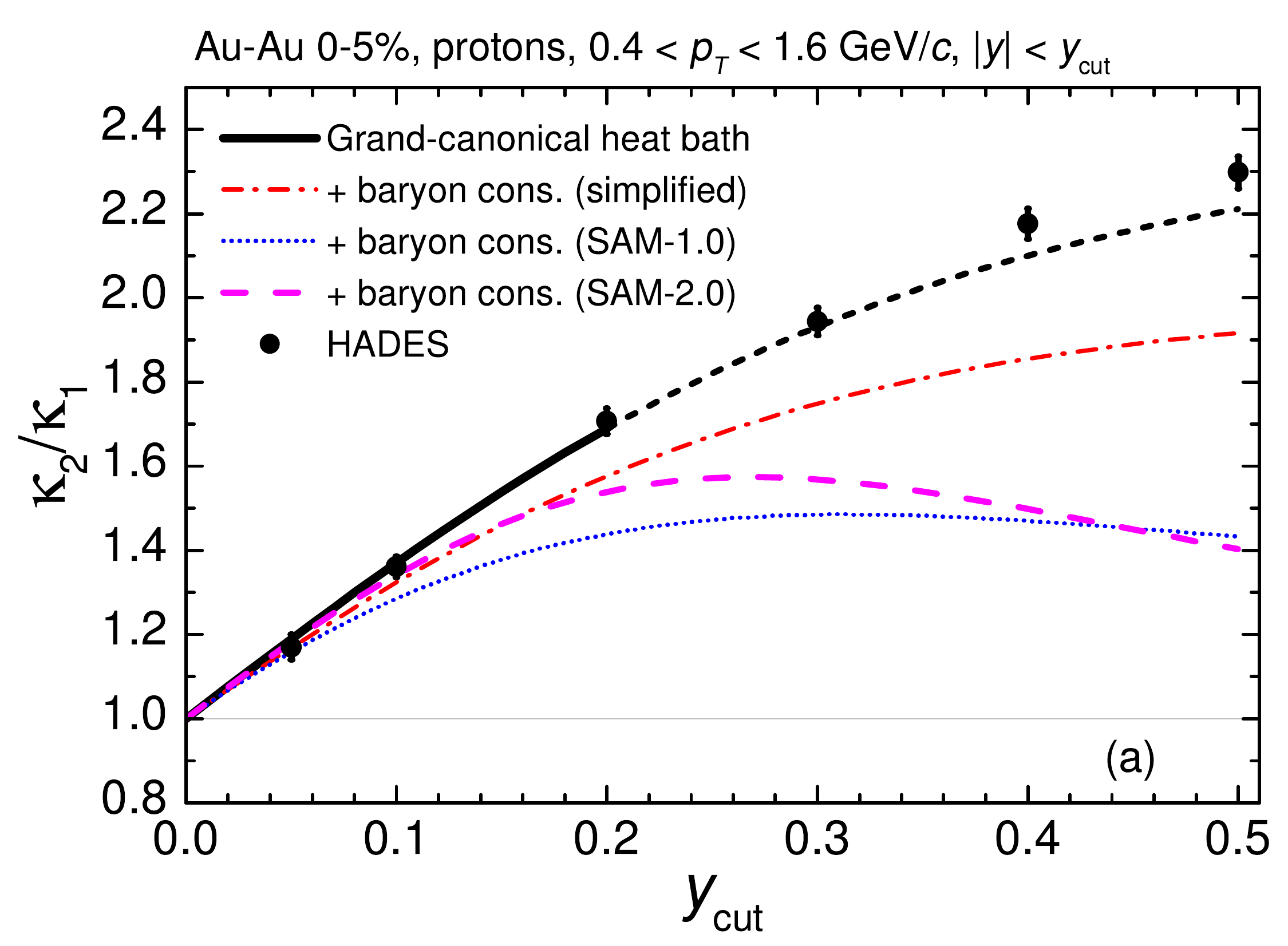}
  \includegraphics[width=.41\textwidth]{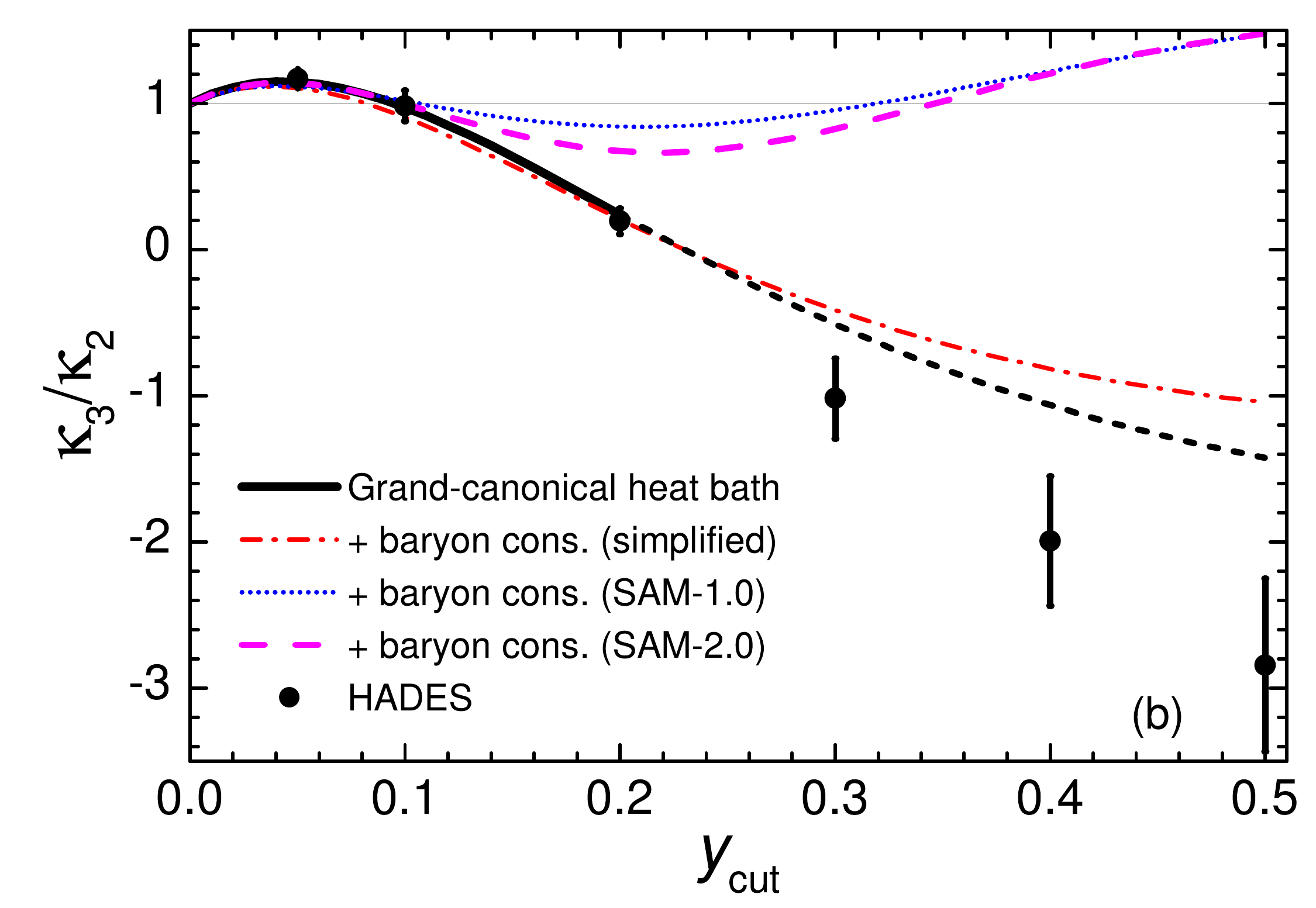}
  \includegraphics[width=.41\textwidth]{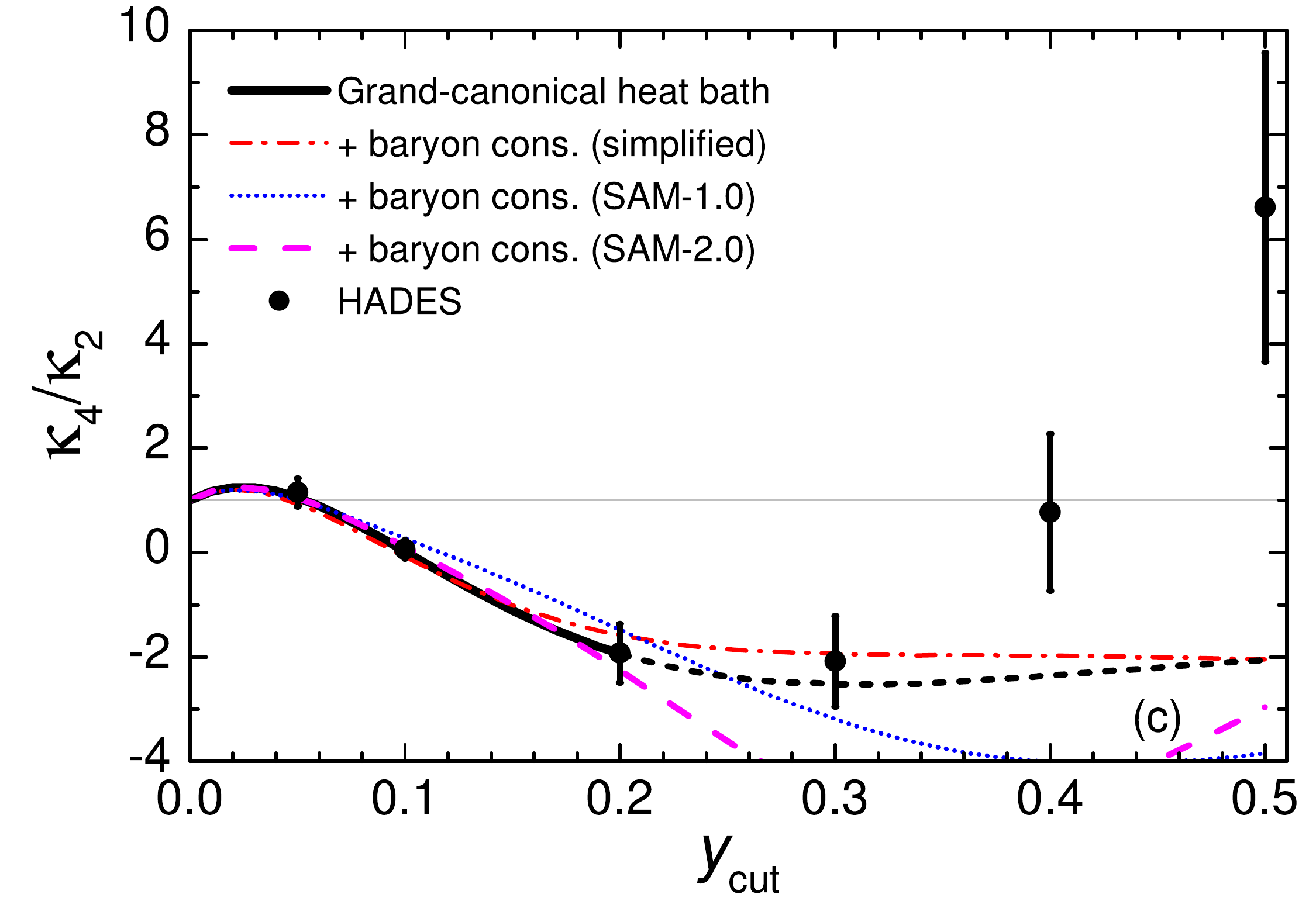}
  \caption{
  Rapidity acceptance dependence of the proton cumulant ratios (a) $\kappa_2/\kappa_1$, (b) $\kappa_3/\kappa_2$, and (c) $\kappa_4/\kappa_2$ in 0-5\% central Au-Au collisions as calculated in the thermodynamic fireball model and compared with the experimental data of the HADES Collaboration~(black circles)~\cite{HADES:2020wpc}.
  Calculations are performed in the grand-canonical limit~(black lines) and with the additional effect of exact baryon conservation implemented via a simplified framework given Eqs.~(\ref{eq:SAMp2}-\ref{eq:SAMp4})~(dash-dotted red lines), SAM-1.0~(dotted blue lines), and SAM-2.0~(dashed magenta lines).
  The dashed parts of the grand-canonical limit calculations correspond to large acceptances where the effect of baryon conservation cannot be neglected.
  }
  \label{fig:K2K4}
\end{figure}

The proton cumulants are evaluated in transverse momentum acceptance $0.4 < p_T < 1.6$~GeV/$c$ around midrapidity, $|y| < y_{\rm cut}$, for various values of $y_{\rm cut}$ consistent with the experimental acceptance~\cite{HADES:2020wpc}.
To fix the values of $\chi_n^B$ we adopt the following strategy: rather than using model assumptions regarding the QCD equation of state at the point of freeze-out, we instead extract the values of $\chi_n^B$ that match the experimental data via Eq.~\eqref{eq:kappanp}.
The values of $\tilde{\alpha}_{n,l}(\Delta p_{\rm acc})$ in Eq.~\eqref{eq:kappanp} are calculated via Eqs.~\eqref{eq:alphatilde},~\eqref{eq:alpha}, and~\eqref{eq:pacc} where $\Delta p_{\rm acc} = \{ 0.4 < p_T < 1.6~\text{GeV}/c,~|y| < y_{\rm cut}\}$. 
We perform this procedure for values of $y_{\rm cut}$ in a range $0.1-0.2$, using central values of the experimentally measured cumulants. 
We do not use the data with $y_{\rm cut} > 0.2$ because at large acceptance the effect of exact baryon conservation becomes dominant~(see the next section below).
One obtains the following grand-canonical baryon susceptibility ratios: $\chi_2^B / \chi_1^B = 8.95$, $\chi_3^B / \chi_2^B = -32.3$, $\chi_4^B / \chi_2^B = 641$ for $y_{\rm cut} = 0.1$ and $\chi_2^B / \chi_1^B = 9.38$, $\chi_3^B / \chi_2^B = -33.9$, $\chi_4^B / \chi_2^B = 741$ for $y_{\rm cut} = 0.2$.
The extracted values are qualitatively consistent between the two data sets corresponding to different $y_{\rm cut}$ and can be combined as
\eq{\label{eq:chiBres}
\frac{\chi_2^B}{\chi_1^B} & \sim 9.17 \pm 0.21, \\
\frac{\chi_3^B}{\chi_2^B} & \sim -33.1 \pm 0.8, \\
\label{eq:chiBresend}
\frac{\chi_4^B}{\chi_2^B} & \sim 691 \pm 50. 
}
The theoretical uncertainty here comes from the differences in the values of the extracted susceptibilities for $y_{\rm cut} = 0.1$ and 0.2.

Then we use the extracted values of $\chi_n^B$ to evaluate the proton cumulants as a function of $y_{\rm cut}$.
Figure~\ref{fig:K2K4} depicts the behavior of the corresponding proton cumulant ratios $\kappa_2^p/\kappa_1^p$, $\kappa_3^p/\kappa_2^p$, and $\kappa_4^p/\kappa_2^p$ in the model using the central values of the susceptibilities from Eqs.~(\ref{eq:chiBres}-\ref{eq:chiBresend}) as input. This is shown by the black lines.
The model calculations are compared with the HADES data.
As a consistency check, the model reproduces the HADES data within the errors for the cumulant ratios measured with $y_{\rm cut} = 0.05$. Note that these data were not used in the procedure to extract the susceptibilities.
Remarkably, the model also reproduces semi-quantitatively the experimental data for $y_{\rm cut} > 0.2$, even though it does not incorporate the exact baryon conservation which is relevant at large acceptances.

Due to the limitations and simplifications of our approach, the extracted baryon number susceptibilities should be taken with a grain of salt from a point of view quantitative estimation of the equilibrium grand-canonical susceptibilities.
Qualitatively, however, it is clear that the results indicate the presence of large non-Gaussian fluctuations of the baryon number, characterized by the increasing magnitude of the high-order cumulants. In fact, the extracted baryon susceptibilities exhibit the following hierarchy
\eq{\label{eq:chiBstruc}
\chi_4^B \gg -\chi_3^B \gg \chi_2^B \gg \chi_1^B.
}

One known physical mechanism that can generate large non-Gaussian fluctuations of the baryon number in the grand-canonical ensemble is the QCD critical point~\cite{Stephanov:2008qz,Stephanov:2011pb}.
If this is the correct mechanism, it would indicate that the freeze-out point of $\sqrt{s_{\rm NN}} = 2.4$~GeV Au-Au collisions is located in the critical scaling region in close proximity to the critical point, where the large non-Gaussian fluctuations are predicted to emerge.\footnote{Specifically, the sign structure of the susceptibilities in Eq.~\eqref{eq:chiBstruc} would indicate the freeze-out of baryon fluctuations on the quark side of the critical point.}
Assuming that the freeze-out indeed takes place close to the critical point, one can use this to estimate its location in the QCD phase diagram.
The values of $T$ and $\mu_B$ at the chemical freeze-out have been estimated by fitting the measured hadron yields in the framework of statistical hadronization model~(SHM) in Refs.~\cite{Harabasz:2020sei,Motornenko:2021nds}. Although these fits show two distinct minima that describe the data similarly well, it was shown in Ref.~\cite{Motornenko:2021nds} using transport model simulations that only one of the two minima corresponds to a reasonable chemical freeze-out picture, and it is located at 
\eq{\label{eq:CPlocation}
T \sim 70~\rm{MeV}, \qquad \mu_B \sim 750-800~\rm{MeV}.
}
As discussed in~\cite{Motornenko:2021nds}, the extracted values are fairly robust with respect to the theoretical uncertainties in the SHM, in particular, to the treatment of nucleon interactions.
Note also, that the extract chemical freeze-out temperature of 70 MeV is the same as the $T_{\rm kin} = 70$~MeV value describing the momentum distributions of protons and pions in the fireball model that we used, consistent with the notion of approximately simultaneous chemical and kinetic freeze-out. 
If the interpretation of the experimental data presented here is correct, the thermal freeze-out at HADES is located close to the QCD critical point and thus
Eq.~\eqref{eq:CPlocation} serves as an estimate for the possible location of the QCD critical point.
The estimate, which corresponds to $\mu_B/T \sim 11$, puts the critical point beyond the reach of current lattice QCD methods corresponding to $\mu_B/T \lesssim 2-3.5$~\cite{HotQCD:2018pds,Borsanyi:2021sxv} but in the ballpark of predictions of some effective QCD models like those based on holography~\cite{Critelli:2017oub,Grefa:2021qvt}.

The extracted susceptibilities can also be used to improve the analysis presented in Ref.~\cite{Sorensen:2021zme}, where it was shown that the cumulants can be used to estimate the (isothermal) speed of sound, as well as its logarithmic derivative with respect to baryon density.
To do that one needs to determine the true grand-canonical cumulants of baryon number whereas in Ref.~\cite{Sorensen:2021zme} direct measurements of the momentum space proton number cumulants were used as a proxy instead.
Using Eqs.~\eqref{eq:chiBres} and~\eqref{eq:CPlocation}, we obtain the following for the speed of sound:
\eq{\label{eq:cT}
c_T^2 \approx \frac{T \, \chi_1^B}{\mu_B \, \chi_2^B} \sim (9.8 \pm 0.2) \cdot 10^{-3},
}
which is a small value consistent with proximity to the critical point.
The logarithmic derivative reads
\eq{\label{eq:dlogcT}
\left( \frac{d \ln c_T^2}{d \ln n_B} \right)_T & = \frac{n_B}{c_T^2} \, \left( \frac{d c_T^2}{d n_B} \right)_T \nonumber \\ & \approx 1 - \frac{\chi_3^B \, \chi_1^B}{(\chi_2^B)^2} - c_T^2 \sim 4.6 \pm 0.1~.
}
Our results agree with the simplified estimates of Ref.~\cite{Sorensen:2021zme} qualitatively, while quantitatively they indicate a smaller $c_T^2$ and a larger $d (\ln c_T^2) / d (\ln n_B)$ at the HADES collision energy than originally obtained in \cite{Sorensen:2021zme}.

\paragraph{\bf Effect of exact baryon number conservation.}

Our analysis has so far been based on the picture of thermal particle emission from a grand-canonical heat bath.
This does neglect one important effect, namely the exact conservation of the total baryon number, which does not change throughout the heavy-ion collision~\cite{Bzdak:2012an}.
In addition, other exact conservation laws like that of electric charge or total energy-momentum can influence the results as well.
Correcting the results for conservation laws is challenging, especially at lower collision energies, although theoretical advances have been made recently~\cite{Vovchenko:2020tsr,Vovchenko:2020gne}.
Here, we use two different methods to incorporate baryon conservation in order to estimate the systematic error due to this effect.
First, we consider a simplified correction based on Refs.~\cite{Bzdak:2012an,Vovchenko:2020tsr} that originally has been derived within the ideal gas model in the canonical ensemble and discussed in the original HADES paper~\cite{HADES:2020wpc}. The proton cumulant ratios subject to baryon conservation are computed in this simplified framework as
\eq{\label{eq:SAMp2}
\frac{\tilde{\kappa}_2^p}{\tilde{\kappa}_1^p} & = (1-\alpha) \, \frac{\kappa_2^{p}}{\kappa_1^{p}}, \\
\label{eq:SAMp3}
\frac{\tilde{\kappa}_3^p}{\tilde{\kappa}_2^p} & = (1-2\alpha) \, \frac{\kappa_3^{p}}{\kappa_2^{p}}, \\
\label{eq:SAMp4}
\frac{\tilde \kappa_4^p}{\tilde \kappa_2^p} & = (1-3\alpha \beta) \, \frac{\kappa_4^{p}}{\kappa_2^{p}} - 3\alpha \beta \left(\frac{\kappa_3^{p}}{\kappa_2^{p}}\right)^2. 
}
We denote by tilde the cumulants that are subject to the effect of exact baryon conservation.
Here $\beta = 1 - \alpha$ and $\alpha = \mean{N_p^{\rm acc}}/\mean{N_B^{\rm tot}}$ is the fraction of measured protons from all baryons. This fraction varies from $\alpha \approx 0.03$ for $y_{\rm cut} = 0.1$ to $\alpha \approx 0.13$ for $y_{\rm cut} = 0.5$.
The results for proton cumulant ratios corrected for baryon conservation in this way are shown by the dash-dotted lines in Fig.~\ref{fig:K2K4}.
These results show the same qualitative $y_{\rm cut}$ dependence as the uncorrected results, although they do worsen the quantitative agreement with the experimental data, especially for the scaled variance.
It is possible that the agreement could be recovered by readjusting the values of the susceptibilities.

However, it should be noted that the presented correction of proton cumulants for baryon conservation cannot be expected to be accurate. 
In particular, using the fraction of accepted protons, i.e. excluding other baryons such as neutrons and light nuclei, for calculating $\alpha$ is only justified in the limit of free gas, where the grand-canonical susceptibilities $\chi_n^B$ correspond to Poisson statistics, i.e. if the susceptibilities of all orders are equal. The values of the extracted susceptibilities in Eqs.~(\ref{eq:chiBres}-\ref{eq:chiBresend}) clearly violate this assumption. In this case, the corrections in Eqs.~(\ref{eq:SAMp2}-\ref{eq:SAMp4}) should only be applied to cumulants of the conserved baryon number rather than that of non-conserved proton number, as discussed in Ref.~\cite{Vovchenko:2020tsr} in the framework of the subensemble acceptance method~(SAM-1.0).
Therefore, one should apply the correction to the cumulants of accepted \emph{baryons} in Eq.~\eqref{eq:kappaBacc}~(rather than protons) using the following expressions from the SAM-1.0:
\eq{\label{eq:SAMB1}
\tilde{\kappa}_1^B & = \kappa_1^{B}, \\
\label{eq:SAMB2}
\tilde{\kappa}_2^B & = (1-\alpha) \, \kappa_2^{B}, \\
\label{eq:SAMB3}
\tilde{\kappa}_3^B & = (1-2\alpha) \, \kappa_3^{B}, \\
\label{eq:SAMB4}
\tilde \kappa_4^B & = (1-3\alpha \beta) \, \kappa_4^{B} - 3\alpha \beta \frac{\left(\kappa_3^{B}\right)^2}{\kappa_2^{B}}. 
}
The cumulants $\tilde \kappa_n^p$ of accepted protons are then obtained from $\tilde \kappa_n^B$ by convoluting the latter with the binomial distribution with acceptance $q$, as before.
The results of this procedure are shown in Fig.~\ref{fig:K2K4} by the dotted lines.
In addition to SAM-1.0, we also apply an updated framework called SAM-2.0, which has recently been developed in Ref.~\cite{Vovchenko:2021yen} and used to describe experimental measurements of proton cumulants in Au-Au collisions for $\sNN \gtrsim 20$~GeV.
The advantage of SAM-2.0 over SAM-1.0 is that its formulation is not restricted to a uniform coordinate space but allows one to perform the corrections in momentum space, which need not be homogeneous.
The method involves calculating the joint cumulants proton/baryon distribution both inside and outside the acceptance, and then using a mapping function to evaluate the cumulants of accepted protons subject to baryon conservation. The details of the mapping function and the evaluation procedure for an emission from a Cooper-Frye hypersurface can be found in Refs.~\cite{Vovchenko:2021yen} and~\cite{Vovchenko:2021kxx}, respectively.
The results obtained within SAM-2.0 are shown in Fig.~\ref{fig:K2K4} by the dashed lines.

The results obtained using SAM-1.0 and SAM-2.0 are similar.
It is evident that the effect of baryon conservation computed by either of the two methods is strong for $y_{\rm cut} > 0.2$, where the model disagrees with the experimental data. 
We checked within SAM-2.0 that the agreement cannot be recovered by refitting the values of $\chi_n^B$: one can fit $\chi_n^B$ to describe the data for a particular value of $y_{\rm cut} > 0.2$, however, the model then fails to describe the data for other rapidity cuts.
The SAM-2.0 fits for  $y_{\rm cut} = 0.1$ and $y_{\rm cut} = 0.2$, however, reproduce the hierarchy~\eqref{eq:chiBstruc} given by Eqs.~\eqref{eq:chiBres}-\eqref{eq:chiBresend}, indicating that corrections due to baryon number conservation are indeed subleading for $y_{\rm cut} \lesssim 0.2$.

The results indicate that a more involved modeling is needed to describe the cumulants at $y_{\rm cut} > 0.2$.
On one hand, as discussed in Ref.~\cite{Vovchenko:2021yen}, the applicability conditions of the SAM at low collision energies may not be satisfied well, and thus a more involved method for estimating the effect baryon conservation may be needed.
Furthermore, the additional effect of exact conservation of electric charge can be relevant as well~\cite{Vovchenko:2020gne}.
Thus, more involved modeling may be warranted than the fireball model considered here. One possibility is to use molecular dynamics with a critical point as recently explored in Ref.~\cite{Kuznietsov:2022pcn}.
On the other hand, it is also possible that the effects of spectator-participant interactions become sizable at larger $y_{\rm cut}$, in particular volume fluctuations, thus invalidating the thermodynamic approach for proton cumulants at large rapidities.
It is also evident from the results that the baryon conservation effect is moderate and can likely be neglected for $y_{\rm cut} < 0.2$. The data analysis in that regime can thus be performed in the grand-canonical ensemble, without the theoretically challenging modeling of exact baryon conservation.

\paragraph{\bf Summary.}

We have shown how experimentally measurable proton number cumulants in heavy-ion collisions at high $\mu_B$ can be related to baryon number susceptibilities in the picture of thermal emission of particles from  approximately uniform fireball superimposed on hydrodynamic flow velocity profile.
We applied the formalism to recent experimental data of the HADES Collaboration and showed that the acceptance dependence of proton cumulants at $|y_{\rm cut}| \lesssim 0.2$ in Au-Au collisions at $\sNN = 2.4$~GeV is consistent with a thermal emission of nucleons from a grand-canonical heat bath. 
The extracted baryon number susceptibilities~[Eqs.~(\ref{eq:chiBres}-\ref{eq:chiBresend})] exhibit an hierarchy 
$\chi_4^B \gg -\chi_3^B \gg \chi_2^B \gg \chi_1^B$,
indicating large non-Gaussian fluctuations that, naively, could point to the presence of the QCD critical point close to the chemical freeze-out of $\sqrt{s_{\rm NN}} = 2.4$~GeV Au-Au collisions at $T \sim 70$~MeV and $\mu_B \sim 850-900$~MeV.
However, the description of the experimental data at higher rapidity acceptances $y_{\rm cut} > 0.2$ becomes challenging once the effect of exact baryon number conservation is incorporated. 
This leads to a conundrum: the experimental data are better described at $y_{\rm cut} > 0.2$ by thermodynamic model without exact baryon conservation compared to the thermodynamic model incorporating this necessary ingredient.
The following interpretations (or combinations thereof) of the obtained results are possible, which shall be clarified by future studies and experimental measurements:
\begin{enumerate}
    \item The thermodynamic approach to proton emission is not the correct mechanism for their production at HADES energies, in which case the present approach cannot be used to reliably extract the baryon susceptibilities.
    The fact that the data do not follow the trend which would typically be expected from baryon conservation should still be understood regardless of the fireball model, especially since baryon conservation appears to be driving the behavior seen in the proton cumulant measurements at $\sNN \gtrsim 20$~GeV~\cite{Vovchenko:2021kxx}.
    \item The thermodynamic approach is the correct mechanism~(at least for the participant matter) but the effect of exact baryon conservation incorporated through the SAM is not accurate, given the limitations of the SAM at lower collision energies discussed in Refs.~\cite{Vovchenko:2020tsr,Vovchenko:2021yen}. 
    In this case, the estimates given by Eqs.~(\ref{eq:chiBres}-\ref{eq:chiBresend}) may be accurate since they were obtained using the data at $y_{\rm cut} < 0.2$ where the effect of exact baryon conservation is subleading. 
    This interpretation can be tested by developing a more accurate treatment of baryon conservation~(as well as electric charge and strangeness conservation) at HADES energies.
    \item The experimental data are driven by non-dynamical contributions to proton fluctuations which should be removed before meaningful comparisons to the thermodynamic model can be made. Note that one such source of non-dynamical fluctuations, namely the participant (or volume) fluctuations, have been removed from the data as detailed in Ref.~\cite{HADES:2020wpc}.
\end{enumerate}

We hope that future theoretical and experimental studies will allow one to reach a firm conclusion and that the method developed in the present work can be used in such studies.

\begin{acknowledgments}

\emph{Acknowledgments.} 
This work received support through the U.S. Department of Energy, 
Office of Science, Office of Nuclear Physics, under contract number 
DE-AC02-05CH11231231 and within the framework of the
Beam Energy Scan Theory (BEST) Topical Collaboration.
V.V. acknowledges the support through the
Feodor Lynen Program of the Alexander von Humboldt
foundation. 
\end{acknowledgments}

\bibliography{HADES-fluctuations}

\end{document}